\documentclass[a4paper, 11pt]{article}
\usepackage[T1]{fontenc}
\usepackage[utf8]{inputenc}
\usepackage[english]{babel}
\usepackage{amsmath}
\usepackage{amssymb}
\usepackage{braket}
\usepackage{mathtools}
\usepackage{dsfont}
\usepackage{subcaption}
\usepackage{array}
\usepackage{booktabs}
\usepackage{yfonts}
\usepackage{setspace}
\usepackage{verbatim}

\usepackage[usenames,dvipsnames]{color}

\numberwithin{equation}{section}

\def\be{\begin{equation}}
\def\ee{\end{equation}}

\newcommand{\nn}{\nonumber}
\newcommand{\diff}{\mathrm{d}}

\definecolor{applegreen}{rgb}{0.55, 0.71, 0.0}

\usepackage{jheppub}                   

\makeatletter
\gdef\@fpheader{\ }                    
\makeatother

\title{Non-spinning tops are stable}

\author[a]
{Iosif Bena,}
\author[a, b]
{Giorgio Di Russo,}
\author[b]
{Jose Francisco Morales,}
\author[b]
{Alejandro Ruip\'erez,}

\emailAdd{dirusso@roma2.infn.it, iosif.bena@ipht.fr, morales@roma2.infn.it, alejandro.ruiperez@roma2.infn.it}

\affiliation[a]{Institut de Physique Th\'eorique, Universit\'e Paris Saclay, CEA, CNRS, \\
F-91191 Gif-sur-Yvette, France}
\affiliation[b]{Dipartimento di Fisica, Universit\`a di Roma ``Tor Vergata'' \& INFN  Roma 2,\\ Via della Ricerca Scientifica 1, 00133, Roma, Italy}

\abstract{We consider coupled gravitational and electromagnetic perturbations of a family of five-dimensional Einstein-Maxwell solutions that describes both magnetized black strings and horizonless topological stars. We find that the odd perturbations of this background lead to a master equation with five Fuchsian singularities and compute its quasinormal mode spectrum using three independent methods: Leaver, WKB and numerical integration. 
Our analysis confirms that odd perturbations always decay in time, while spherically symmetric even perturbations may exhibit for certain ranges of the magnetic fluxes instabilities of Gregory-Laflamme type for black strings and of Gross-Perry-Yaffe type for topological stars.  This constitutes evidence that topological stars and black strings are classically stable in a finite domain of their parameter space.}

\begin{document}

\allowdisplaybreaks
\maketitle



\section{Introduction}

The physics of black holes is responsible for some of the deepest puzzles in our quest to formulate a unified quantum theory of gravity. On one hand, black holes have an entropy proportional to the area of the event horizon in Planck units, and hence, according to Statistical Mechanics, an enormous number of states ($e^{10^{90}}$ for the Sgr.A black hole at the center of the Milky Way). On the other hand, uniqueness theorems indicate that General Relativity is not able to distinguish any of these states and this, in turn, leads to violations of quantum unitarity. It is also possible to argue  \cite{Mathur:2009hf,Almheiri:2012rt,Guo:2021blh} that the only way to avoid such violations is if these states give different physics from that of the classical General-Relativity black-hole solution at the scale of the event horizon.

However, constructing such states is no easy feat: as the black-hole horizon is null, any horizon-sized object we can construct using normal four-dimensional matter will immediately fall in. The only mechanism to avoid gravitational collapse in a classical theory is to use (or mimic) higher-dimensional theories with nontrivial fluxes wrapping topologically-nontrivial cycles   \cite{Gibbons:2013tqa,deLange:2015gca}, of the type one naturally finds in String Theory. 

Furthermore, the black hole horizon is perhaps the only thing in our universe that grows when gravity becomes stronger. Hence, if one wants to construct horizon-sized extreme-compact-objects (ECO's) that replace the black hole, these objects must contain non-perturbative solitonic degrees of freedom, which become lighter and larger when gravity becomes stronger. 

For supersymmetric and extremal-non-supersymmetric black holes there is an almost-20-year history of string-theory and supergravity constructions of such horizon-sized ECO's\footnote{See \cite{Bena:2022rna} and references therein}, which are also known as microstate geometries, fuzzballs geometries, or ``topological stars" \cite{Bena:2013dka}. The string-theory ingredients that enter their construction have exactly the properties needed to avoid collapse and grow when gravity becomes stronger. However, constructing and analyzing topological stars for non-supersymmetric black holes is much more challenging, requiring in general solving non-linear PDE's that, absent supersymmetry, do not factorize. 

Besides some artisanal solutions \cite{Jejjala:2005yu, Bena:2009qv}, there are now three systematic routes to build such solutions. The oldest is the floating-JMaRT factorization \cite{Bossard:2014ola, Bossard:2014yta}, which can in principle produce non-extremal rotating solutions, but which unfortunately does not seem to produce solutions that have the same charges and angular momenta as non-extremal black holes with a macroscopic horizon \cite{Bena:2015drs,Bena:2016dbw, Bossard:2017vii}. The second is to use numerics in a consistent truncation to three-dimensional supergravity \cite{Mayerson:2020tcl} to produce {\em microstrata} \cite{Ganchev:2021pgs, Ganchev:2021ewa,Ganchev:2023sth,Houppe:2024hyj} that have the same charges as non-extremal asymtptotically-AdS rotating black holes. The third is to use the Bah-Heidmann factorization \cite{Bah:2020pdz} to produce non-extremal non-rotating multi-bubble topological stars, which can have both flat-space and AdS asymptotics \cite{Bah:2020ogh,Bah:2021owp,Bah:2021rki, Bah:2021irr,Heidmann:2021cms,Bah:2022pdn,Bah:2022yji,Bah:2023ows}. Absence of rotation notwithstanding, this later method appears to be the most prolific at generating solitonic solutions with non-extremal black-hole charges and mass, including Schwarzschild microstate \cite{Bah:2022yji,Bah:2023ows}.

The main question about these non-extremal microstate geometries is whether they are stable or unstable, and what are the physical implications of their stability or absence thereof. Since the asymptotically-AdS$_3$ solutions are dual to coherent CFT states that have both right movers and left movers, one can argue that generic solutions with a large number of identical CFT strands should be unstable \cite{Chowdhury:2008bd}, much like the JMaRT solution is  \cite{Cardoso:2005gj, Chowdhury:2007jx, Bianchi:2023rlt}. However, this argument does not extend to solutions without an AdS near-horizon region, such as Schwarzschild microstates. 

The purpose of this paper is to investigate the stability under coupled gravitational and electromagnetic perturbations of the simplest spherically-symmetric topological star, which can be constructed as four-dimensional Euclidean Schwarzschild (ES) solution times time, to which one adds magnetic flux on the ES bolt.  The solution is specified by two parameters  $r_s$ and $r_b$ that interpolate between a magnetic black string solution with a finite-area event horizon when $r_s> r_b$  and a smooth horizonless solution one when $r_s< r_b$.  Throughout this paper we will call the former {\em the black string} and the later  {\em the top star}.
 The top star solution is one of the key ingredients in the building of the more general multi-bubble Bah-Heidmann solutions, and hence our investigation is the first step in a programme to determine the stability or instability of these solutions and the physical consequences thereof\footnote{The stability of top stars under scalar perturbations was established in \cite{Bianchi:2023sfs, Heidmann:2023ojf, Cipriani:2024ygw}.}.

  The equations governing the perturbation around the black string and top star are the same, and couple the electromagnetic fluctuations to the gravitational ones. They separate into a rather intractable parity-even sector and a more tractable parity-odd sector, on which we focus in this paper. We find that in this sector the perturbations separate into two coupled systems of ODE's.  We show that one of these systems boils down to two independent second-order ODE's for the so-called \emph{master variables}. These ODE's have five Fuchsian singularities, and cannot be mapped to a Heun equation or confluent version thereof.  Hence, they are harder to solve then the scalar perturbations of black holes \cite{Regge:1957td,Teukolsky:1973ha}, of D-brane bound states \cite{Bianchi:2021mft,Bianchi:2021mft}, and of topological stars \cite{Bianchi:2023sfs, Heidmann:2023ojf, Cipriani:2024ygw}, where the linear dynamics is described by confluent forms of the Heun equation.

 We primarily rely on the Leaver method to find the quasinormal (QNM) frequencies of topological stars and magnetic black strings. The then verify the results against those obtained through direct numerical integration and WKB analysis. The WKB approximation works well when the orbital number of the perturbation is large. Direct numerical integration of the differential equation produces reliable results for frequencies with a small imaginary part. Our calculations show that QNM frequencies of odd perturbations have always negative imaginary part, so they represent modes decaying in time. 

For even perturbations, we are able to separate the equations only for spherically symmetric perturbations and find stable  solutions for all topological stars with $r_s < 2 r_b$ and black strings with $r_b< 2 r_s$ in agreement with \cite{Bah:2021irr,Stotyn:2011tv}. 
Our analysis provide a unifying picture that interpolates between the Gross-Perry-Yaffe-type instability of top stars with zero or small magnetic fields \cite{Gross:1982cv,Witten:1981gj,Bah:2021irr,Stotyn:2011tv} and the Gregory-Laflamme instability \cite{Gregory:1993vy,Gregory:1994bj} of magnetized black strings with zero or small magnetic charge \cite{Miyamoto:2007mh}. Both these instabilities manifest themselves through the existence of QNMs that blow up in time.

The key result of our calculation is that in the regimes of parameters where these instabilities are absent, all other perturbations decay in time. Hence, in these regimes, both the top star and the magnetized black strings are stable!

In Section~\ref{sec:pert} we derive the system of ordinary differential equations governing the odd sector of the coupled gravitational and electromagnetic perturbations.   
In Section~\ref{sec:qnm} we compute the spectrum of QNM frequencies for odd perturbations. In Section~\ref{sec:GL} we determine the regimes of Gregory-Laflamme instability. Finally, in Section~\ref{sec:conclusions} we present some conclusions and future directions.

{\bf Note added:} When this work was nearly complete, we were informed that another group was working independently on the same problem~\cite{Paolo-new}. Although there is significant overlap between our results, our analysis and that of~\cite{Paolo-new} also focus on different aspects and numerical methods and are therefore complementary to each other. We have compared some of our numerical results to those of~\cite{Paolo-new}, finding very good agreement.


\section{Gravitational and electromagnetic perturbations}
\label{sec:pert}

We consider solutions of Einstein-Maxwell theory in five dimensions describing magnetically-charged topological stars and black strings. Our goal is to study linear perturbations of the metric and gauge field around these backgrounds. The study of perturbations around magnetically charged backgrounds poses a technical problem well known in the literature on black-hole perturbation theory \cite{Gerlach:1980tx, Kodama:2003kk, Pereniguez:2023wxf}: the usual decoupling between even and odd perturbations does not occur.  Here we circumvent this problem by dualizing the vector field into a two-form potential $C_{\mu \nu}$ with field strength $F_{\mu\nu\rho}=3 \partial_{[\mu}C_{\nu\rho]}$. In terms of this field, the equations of motion are
\begin{eqnarray}
\label{eq:einstein}
R_{\mu\nu}-\frac{1}{2}g_{\mu\nu}R&=&\frac{1}{4}\left(F_{\mu\alpha\beta}F_{\nu}{}^{\alpha\beta}-\frac{1}{6}g_{\mu\nu}F_{\alpha\beta\gamma}F^{\alpha\beta\gamma}\right)\,,\\[1mm]
\label{eq:Maxwell}
\nabla_{\mu}F^{\mu\nu\rho}&=&0\,. 
\end{eqnarray}


We study linear perturbations around solutions of \eqref{eq:einstein} and \eqref{eq:Maxwell} with metric ${\bar g}_{\mu\nu}$ and two-form potential ${\bar C}_{\mu\nu}$ given by \cite{Horowitz:1991cd, Bah:2020ogh, Bah:2020pdz} 
\begin{eqnarray}
\diff {\bar s}^2&=&-f_s(r)\diff t^2+\frac{\diff r^2}{f_s(r) f_b(r)}+r^2 \left(\diff \theta^2 +\sin \theta^2 \diff \phi^2\right)+f_b(r)\diff y^2\,,\\[1mm]
{\bar C}&=&\frac{\sqrt{3r_s r_b}}{r}\,\diff t\wedge  \diff y\, ,
\end{eqnarray}
where $r_s$, $r_b$ are some real positive numbers and
\begin{equation}
f_{s}(r)=1-\frac{r_{s}}{r}\, , \hspace{1cm} f_{b}(r)=1-\frac{r_{b}}{r}\, .
\end{equation}
There are two regimes of interest:
\begin{eqnarray}
{\rm Black~string}:  && \qquad  r_b<r_s \nn\\
{\rm Topological~star}:  && \qquad  r_s<r_b  
\end{eqnarray}
The first is called the black string regime \cite{Horowitz:1991cd}, in which  the solution has an event horizon at $r=r_s$. In the second, the solution describes a topological star \cite{Bah:2020ogh, Bah:2020pdz}, which is a smooth horizonless solution provided the coordinate $y$ is periodically identified with period
\begin{equation}\label{eq:id_y}
y\sim y+\frac{4\pi r_b^{3/2}}{\sqrt{r_b-r_s}}\, .
\end{equation}
For this periodicity the spacetime smoothly ends at the hypersurface $r=r_b$. The geometry near $r=r_b$ is ${\mathbb R}_t\times {\mathbb R}^2\times {\mathbb S}^2$.

\subsection{Linear perturbations}

 We consider perturbations of the background metric and two-form potentials,
 \begin{equation}
g_{\mu\nu}={\bar g}_{\mu\nu}+h_{\mu\nu}\,, \hspace{1cm} C_{\mu\nu}={\bar C}_{\mu\nu}+c_{\mu\nu}\,.
\end{equation}
Following \cite{Regge:1957td}, we separate the perturbations according to their transformations under parity in even- and odd-types. For the background solutions we are considering, even and odd perturbations do not couple. Therefore, we can study them separately. Here we focus on odd perturbations, which are considerably simpler. These are given by
\begin{eqnarray}
\label{eq:metricpert}
h_{\mu\nu}\diff x^{\mu}\diff x^{\nu}\, &=&\,2 \, e^{-{\rm i} \omega t}  \sum_{a=t,r,y} h_{a}(r) \diff x^a \left[-  \frac{1}{\sin\theta}\frac{\partial Y_{\ell m}}{\partial \phi}\diff \theta+ \sin\theta\frac{\partial Y_{\ell m}}{\partial \theta}\diff \phi\right] \,,\\  
\label{eq:2formpert}
c_{\mu\nu}\diff x^{\mu} \wedge \diff x^{\nu}\, &=&\,2 \, e^{-{\rm i} \omega t}  \sum_{a=t,r,y} c_{a}(r) \diff x^a  \wedge \left[-  \frac{1}{\sin\theta}\frac{\partial Y_{\ell m}}{\partial \phi}\diff \theta+ \sin\theta\frac{\partial Y_{\ell m}}{\partial \theta}\diff \phi\right] \,,
\end{eqnarray}
where $Y_{\ell m}(\theta, \phi)$ are the spherical harmonics satisfying
\begin{equation}
\left[\frac{1}{\sin\theta}\frac{\partial}{\partial\theta}\left(\sin\theta\frac{\partial}{\partial\theta}\right)+\frac{1}{\sin^2\theta}\frac{\partial^2}{\partial\phi^2}+\ell (\ell+1)\right]Y_{\ell m}=0\, .
\end{equation}
 Since the background is spherically symmetric there is no dependence on $m$, so from now on we set $m=0$. In addition, we consider no dependence on the coordinate $y$ for the time being. Plugging \eqref{eq:metricpert} and \eqref{eq:2formpert} into the field equations, \eqref{eq:einstein} and \eqref{eq:Maxwell}, and expanding to linear order in the perturbations, one finds two decoupled systems of ordinary differential equations:  one for
   the  $(h_t,h_r,c_y)$ perturbations
  \begin{eqnarray}
\label{eq:h_0'}
&& h_r'(r)+\left[ \frac{ r(r_s+r_b) -2 r_s r_b}{r (r-r_s)(r-r_b) } \right] h_r(r)+\frac{i r^3 \omega  h_t(r)}{\left(r-r_b\right)
   \left(r-r_s\right){}^2}=0\,,\nn\label{system1} \\[1mm]
   && h_t'(r)-\frac{2 h_t(r)}{r}+{\rm i} \left[\frac{\omega^2 r^3 {-} \left(\ell{+}2\right)\left(\ell{-}1\right)\left(r{-}r_s\right) }{ \omega r^3}\right] h_r(r)  +\frac{\sqrt{3r_b r_s} c_{y}(r)}{ r(r-r_b)}=0\,, \\[1mm]
&& c_{y}'' (r) {+}\frac{r_s c_{y}' (r) }{r \left(r{-}r_s\right)}{+} \left[ \frac{\left(r^5 \omega ^2-\ell (\ell{+}1) r^2 \left(r{-}r_s\right){+}3 r_b r_s \left(r_s{-}r\right)\right)  }{r^2
   \left(r{-}r_b\right) \left(r{-}r_s\right){}^2} \right] c_{y} (r) \nn\\
&& \qquad \qquad\qquad \qquad \qquad \qquad {+}\frac{i \sqrt{3}  (\ell+2) \left(\ell-1\right) \sqrt{r_b r_s} h_r(r) }{r^4
   \omega }=0\nn\,,  
  \end{eqnarray}
  and one for $(c_t,c_r,h_y)$ perturbations
\begin{eqnarray}
\label{eq:h_0'}
&&c_{r}'(r)+ \left[\frac{ r \left(r_b+r_s\right)-2 r_b r_s}{r \left(r-r_b\right) \left(r-r_s\right)} \right] c_{r}(r)+\frac{i r^3 \omega  c_t(r) }{\left(r-r_b\right)
   \left(r-r_s\right){}^2} =0\,, \nn\\[1mm]
&& c_{t}'(r) +\frac{\left[ \ell (\ell+1) (r-r_s)-r^3 \omega ^2\right]c_{r} (r) }{ir^3 \omega }-\frac{\sqrt{3} \ell( \ell+1) \sqrt{r_b r_s}
   h_y(r) }{r \left(r-r_b\right)} =0\,,\label{system2}\\[1mm]
   && h_y'' (r){+}\frac{r_s h_y' (r)}{r(r{-}r_s)}{+} \left[\frac{ \left(r^5 \omega ^2{-}\left(r{-}r_s\right) \left(r
   \left(\ell (\ell{+}1) r{-}2 r_b\right){+}3 r_b r_s\right)\right)}{r^2 \left(r{-}r_b\right) \left(r{-}r_s\right){}^2} \right] h_y(r)  {-}\frac{i \sqrt{3} c_{r}(r) \sqrt{r_b r_s}}{r^4 \omega } =0\nonumber \,.
\end{eqnarray}
Before studying the complete system, we discuss two limits of interest.
 
\subsection{Schwarzschild black string and soliton limits} 

The limits are the $r_b \to 0$ and the $r_s\to 0$ limits. In the first the background solution corresponds to the ``Schwarzschild black string'' (meaning a four-dimensional Schwarzschild solution times a transverse direction). The second limit corresponds to a horizonless solution which we are going to call the Schwarzschild soliton. These solutions are mapped one into another via a double Wick rotation along the coordinates $t$ and $y$ and an exchange of $r_s\leftrightarrow r_b$. This does not imply, as we are going to see, that the equations for the perturbations around these backgrounds coincide (up to an exchange of $r_s \leftrightarrow r_b$). The reason is that here we are considering perturbations with no dependence in $y$. Therefore, the perturbations are not mapped one into each other by the double Wick rotation, even if the background solutions are.
 
 \begin{itemize} 
 
\item \textbf{Schwarzschild black string: $r_b\to0$}.  In this limit the full system reduces to four independent second-order differential equations for $h_r, h_y, c_{r}, c_{y}$. When recast into  Schr\"{o}dinger form, they are given by
\begin{equation}
\Psi_s'' (r) +Q_s(r)\Psi_s (r) =0\,, 
\end{equation}
where $Q_s$ is the spin-$s$ Regge-Wheeler potential \cite{Regge:1957td}
\begin{equation}
\label{eq:odd_pert_schw}
Q_s(r)=\frac{r^4\omega^2-\ell(\ell+1)r(r-r_s)+s^2 r_s \left(r-r_s\right) + \frac{r_s^2}{4}}{r^2 (r-r_s)^2}\,,
\end{equation}
and  
\begin{equation}
\Psi_s\sim\left\{ 
\begin{array}{ccc}
 c_r  & ~~~~  &  s=0  \\
h_y,c_y   &   & s=1   \\
 h_r &   & s=2   
\end{array}
\right.\, .
\end{equation}
 
\item \textbf {Schwarzschild soliton: $r_s\to0$}. The four differential equations can be cast in  Schr\"{o}dinger  form,
\begin{equation}
\Psi_\epsilon''+Q_\epsilon(r)\Psi_\epsilon=0\,, 
\end{equation}
with
\begin{equation}
 Q_\epsilon =\frac{r^3\omega^2-\ell(\ell+1)r + 2\epsilon r_b}{r^2\left(r-r_b\right)}\,,
\end{equation}
and 
\begin{equation}
\Psi_s \sim \left\{ 
\begin{array}{ccc}
c_r,c_y  & ~~~~  &  \epsilon=0  \\
h_r,h_y   &   & \epsilon=1   \\
\end{array}
\right.\, .
\end{equation}
\end{itemize}

\subsubsection{The  $(h_r,c_y,h_t)$ system }

We focus on the system  \eqref{system1} describing the perturbations $(h_t,h_r,c_y)$. The first equation can be solved for $h_t$, leading to a coupled system of two differential equations for $h_r$, $c_y$.  
The system can be decoupled by the linear transformation,
\begin{equation}
\begin{aligned}
h_r =\,& g_h(r) \left[\Psi_+(r)+\Psi_-(r) \right]  \,,\\
c_y =\,&  g_c(r)\left[ (1+\gamma) \Psi_+(r) +(1-\gamma) \Psi_-(r) \right] \,,
\end{aligned}
\end{equation}
with
\begin{equation}
g_h(r)={r^{7\over 2} \over (r-r_b)(r-r_s)^{3/2} }   \, , \hspace{1cm} g_c(r)=-{{\rm i} (2 r_b+3 r_s)   \over 2 \omega  \sqrt{3 r_b r_s} }{r^{1\over 2} \over (r-r_s)^{1\over 2}} \,,
\end{equation}
and
\be
\gamma\equiv \sqrt{1+\frac{12(\ell+2)(\ell-1)r_b r_s}{(2r_b+3r_s)^2}}\,.
\ee
In terms of these variables the system reduces to two decoupled ordinary differential equations of the Schr\"{o}dinger form,
\begin{equation}
\Psi''_{\pm}+Q_{\pm}\Psi_{\pm}=0\,, \label{eqcan}
\end{equation}
with 
\begin{equation}
\begin{aligned}
Q_{\pm}\,=\,&\frac{r^3}{\left(r-r_s\right)^2\left(r-r_b\right)}\left[\omega^2-\frac{\ell(\ell+1)}{r^2}+\frac{\left(2r_b+3r_s\right)\left(1\pm \gamma\right)+2r_s\left( \ell^2+\ell+1\right)}{2r^3}\right.\\[1.5mm]
&\left. -\frac{r_s\left[\left(2r_b+3r_s\right)\left(1\pm \gamma\right)+8r_b+\frac{3}{2}r_s\right]}{2r^4}+\frac{15r_br_s^2}{4r^5}\right]\,. \label{qqtop}
\end{aligned}
\end{equation}
Let us remark that \eqref{eqcan} is a differential equation with five Fuchsian singularities: three regular ones at $r=0,r_s,r_b$, and two colliding at $r=\infty$. Crucially, this radial equation cannot be mapped to a Heun equation or any of its confluent versions, as is the case for the scalar perturbations in the topological star geometry, which are described by a Confluent Heun equation \cite{Bianchi:2023sfs}, similar to all black holes in the Kerr-Newman family \cite{Bianchi:2022wku}.

\subsubsection{The $(h_y, c_t, c_r)$ system}

We can proceed analogously for the system \eqref{system2}. We solve the first equation for $c_t$, which yields a system of two coupled second-order ODEs. Then we look for a change of variables of the form
\begin{equation}
\begin{bmatrix}
h_y\\
 c_{r}\\
\end{bmatrix}=
\begin{bmatrix}
    g_{1+}(r)       & g_{1-}(r)  \\
    g_{2+}(r)       & g_{2-}(r) \\
\end{bmatrix}
\begin{bmatrix}
 \Phi_+\\
\Phi_-\\
\end{bmatrix}\,,
\end{equation}
and fix the functions $g_{1+}, g_{2-}, g_{2+}$ and $g_{2-}$  such that the system decouples in two independent second-order ODEs for $\Phi_{+}$ and $\Phi_{-}$. We find that for this system this not possible. However, for the choice of functions
\begin{equation}
g_{1+}=g_{2+}=1\,, \hspace{1cm} g_{1-}=-g_{2-}=\frac{r^3}{\left(r-r_b\right)\left(r-r_s\right)}\,,
\end{equation}
we find that the second system \eqref{system2} reduces to the following differential equations
\begin{equation}
\begin{aligned}
&\Phi''_{\pm}+\frac{r_s}{r\left(r-r_s\right)} \Phi'_{\pm}+ \frac{[-4 r_b r_s +r\left(2r_b+r_s\right)]\omega \pm {\rm{i}} \sqrt{3 r_b r_s}\,r\left[1+\ell\left(\ell+1
\right)\omega^2\right]}{2r^2\left(r-r_b\right)\left(r-r_s\right)}\Phi_{\mp} \\[1mm]
&+ \left[\frac{r^3\omega^2}{\left(r-r_b\right)\left(r-r_s\right)^2} \pm \frac{{\rm{i}} \sqrt{3 r_b r_s} \,}{2r\left(r-r_b\right)\left(r-r_s\right)}\left(\ell (\ell+1)\,\omega+\frac{1}{\omega}\right)\right.\\[1mm]
&\left.+\frac{2r\left(r_b-\ell (1+\ell)r\right)-r_s\left(r+2r_b\right)}{2r^2\left(r-r_b\right)\left(r-r_s\right)}\right]\Phi_{\pm}=0\,.
\end{aligned}
\end{equation}
Then we can use the equation which does not contain derivatives of $\Phi_{-}$ (equivalently, $\Phi_+$) and solve it (for $\Phi_{-}$). After plugging the resulting expression in the remaining equation, we get a fourth-order ODE for $\Phi_+$. Given its complexity, we omit the details and the study of the QNMs. 


\section{QNMs of topological stars and black strings}
\label{sec:qnm}
 In this section we compute the QNM spectrum associated to the $(h_t,h_r,c_y)$ perturbations, whose dynamics is described by the master equations \eqref{eqcan}. We will mainly use a semi-analytical method developed by Leaver \cite{Leaver:1985ax}, which has been recently applied to the study of scalar perturbations of topological stars, \cite{Bianchi:2023sfs, Heidmann:2023ojf, Cipriani:2024ygw}. Nevertheless, we will compare the results obtained using Leaver's method against those obtained using the WKB approximation and via a direct numerical integration of the differential equation.

 QNMs correspond to solutions of the Schrodinger-type equation (\ref{eqcan}) satisfying the boundary conditions\footnote{We omit the $\pm$ subindex in the master variables for the sake of clarity.}
  \begin{equation}
\Psi(r) \underset{r\to r_0}{\sim}   (r-r_0)^{\lambda_0}\,, \hspace{1cm}   \Psi(r) \underset{r\to \infty}{\sim} r^{\lambda_\infty}  e^{i\omega r}\,. \label{boundary}
\end{equation}
 with
 \be
 r_0=
\left\{
\begin{array}{ccc}
 r_b & ~~~~~  &  {\rm (top~star)}  \\
 r_s  &   &    {\rm (black~string)}\\
\end{array}
\right.
\,, \hspace{2cm}
 \lambda_0 \in
\left\{
\begin{array}{ccc}
 \mathbb{R}_+ & ~~~~~  &  {\rm (top~star)}  \\
 {\rm i} \,  \mathbb{R}_-  &   &    {\rm (black~string)}\\
\end{array}
\right.\, .
 \ee
Namely, we are imposing outgoing boundary conditions at infinity for both background solutions. The difference lies in the boundary conditions imposed at $r_0$. For top stars we just demand regularity at the \emph{cap}: $r=r_b$.  Instead, for the black string we impose incoming boundary conditions at the horizon: $r=r_s$. The specific values of $\lambda_0$ and $\lambda_\infty$ can be obtained by solving the differential equation around $r=r_0$ and $r=\infty$.  The details are given in sections~\ref{sec:qnm_ts} and \ref{sec:qnm_bs}.

Solutions satisfying these boundary conditions exist only for a discrete choice of complex frequencies $\omega_n$: the QNMs. The rest of the section is organized as follows. First, in sections~\ref{sec:wkb} and \ref{sec:num_int} we briefly describe the WKB approximation and the numerical integration methods we are going to use to further confirm the results obtained using Leaver's method. The latter will be discussed in sections~\ref{sec:qnm_ts} (top star) and \ref{sec:qnm_bs} (black string).

\subsection{WKB  approximation}
\label{sec:wkb}

 A rough estimate of the QNM frequencies, $\omega$, can be obtained from a WKB semiclassical approximation of the wave solution \cite{Schutz:1985km, Cardoso:2008bp,Bianchi:2020des,Bianchi:2020yzr,Bianchi:2021mft} around the ``light rings'':  extrema of the effective potential $-Q(r)$ where both $Q$ and its first derivative vanish:
\be
Q(r_c;\omega_c)=Q'(r_c;\omega_c)=0\, .
\ee
To estimate the QNM frequencies, we promote  $\omega_c$ to a complex number (adding a small imaginary part) and demand that the integral between two zeros $r_\pm$  of $Q(r)$ satisfies the Bohr-Sommerfeld quantization condition:
\begin{equation}
\int_{r_-}^{r_+}\sqrt{Q(r;\omega)}dr=\pi\left(n+\frac{1}{2}\right)\, .
\end{equation}
 The solution to linear order in the imaginary part of $\omega$ is given by the WKB formula, namely
 \be
 \omega_{n}^{\rm WKB}=\omega_c   -{\rm i}\left(n+\frac{1}{2}\right) \frac{ \sqrt{ \partial_r^2 Q}}{ \partial_\omega Q}\Big|_{\omega=\omega_c,r=r_c}\,.
 \ee
 
\subsection{Direct numerical integration} 
\label{sec:num_int}

 QNM frequencies can be alternatively obtained via a direct numerical integration of the differential equation in the domain $r\in [r_0,\infty]$ with boundary conditions (\ref{boundary}).  Since both $r=r_0$ and $r=\infty$ are singular points, boundary conditions cannot be imposed exactly at these points, but one can choose some arbitrary points nearby. The procedure has several steps: First, one determines the solution (with the appropriate boundary conditions) as a Taylor expansion around $r_0$. Then, one evaluates the solution at a point $r_0+\epsilon$ with $\epsilon$ small, and uses the result as a boundary condition to find a numerical solution, $\Psi_0$. One then proceeds in the same way to compute a numerical solution $\Psi_{\infty}$  satisfying outgoing boundary conditions at infinity.     
 
The QNM frequencies are then obtained by requiring the matching of the two functions at an intermediate point. This is equivalent to demanding the vanishing of the Wronskian:
 \be
 \Psi_0'(r;\omega) \Psi_{\infty}(r;\omega)- \Psi_0(r;\omega) \Psi_{\infty}'(r;\omega)=0\, , \label{wronskian}
 \ee
 at some point $r$ in the middle. We notice that the Wronskian is independent of the choice of this point. Thus, \eqref{wronskian} becomes an equation for $\omega$ whose solutions are the QNMs. When comparing this method against the Leaver method, we will bear in mind that this numerical method is known to be numerically stable only for frequencies with a small imaginary part \cite{ae16f484-8e04-30db-a6a3-902e6f6fd388, Berti:2009wx, Cardoso:2014sna, Bianchi:2023sfs}. 
  
\subsection{QNMs: top stars}
\label{sec:qnm_ts}
 
 In this section we compute the QNM spectrum of top stars using Leaver's method. As explained in the previous section, top stars correspond to smooth horizonless geometries ending a cap at  $r=r_b$ with $r_b> r_s$. 
Hence, QNMs in this geometry correspond to solutions of \eqref{eqcan} satisfying regular boundary conditions at the cap and behaving as an outgoing wave at infinity.  
Given this, we consider the following ansatz
\begin{equation}\label{rleaver}
\Psi(r)=e^{i\omega r} r^{-{3\over 2} }  (r-r_s)^{\lambda_s}  \sum_{n=0}^\infty
c_n\left(\frac{r-r_b}{r-r_s}\right)^n \, ,
\end{equation} 
with
\be
\lambda_s=\frac{1}{2}+{{\rm i} \omega \over 2} (r_b+2 r_s)\, .
\ee
The expansion (\ref{rleaver}) satisfies the required boundary conditions at the cap and infinity, and solves (\ref{eqcan}) near $r=r_b$ and near $r\to \infty$. 
Plugging (\ref{rleaver}) into (\ref{eqcan}) yields a four-term recursion relation, 
\begin{equation}
\label{rec}
\alpha_n c_{n+1}+\beta_n c_n+\gamma_n c_{n-1}+\delta_n c_{n-2}=0\,, \hspace{5mm} \quad n\geq 0 \,.
\end{equation}
The explicit expressions of the coefficients $\alpha_n, \beta_n, \gamma_n$ and $\delta_n$ for $\Psi_+$ are 
\begin{equation}
\begin{aligned}
\alpha _n \,=\,&  -(n+1) (n+2)\,,\\[1mm]
\beta _n \,=\, & \ell (\ell{+}1)  {+}2 n(n{+}2) {+}1{-} \gamma    +\frac{r_s(3(1+\gamma)-2n(n-2))}{2 r_b} {-}3{\rm  i} (n{+}1)
   \omega  r_b{-}\frac{\omega ^2 r_b^3}{r_b{-}r_s}\,,\\[1mm]
   \gamma _n \,=\,&{-}n (n{+}1){+}1{+}\gamma {+} \frac{r_s (3 \gamma{-}4 (n{-}3) n{-}2 \ell  (\ell {+}1){-}5)}{2 r_b}+\frac{1}{2} i \omega 
   \left(2 n r_b{+}r_b{+}(10 n{-}7) r_s\right){+}\frac{\omega ^2 r_b^2 \left(r_b{+}7
   r_s\right)}{4 \left(r_b{-}r_s\right)}\,,\\[1mm]
   \delta _n \,=\,&  \frac{(n-2) r_s \left(n-2 i \omega  r_s-2\right)}{r_b}-\frac{1}{4} \omega  r_s
   \left(\omega  r_b+4 i (n-2)+4 \omega  r_s\right)-\frac{\omega ^2
   r_s^3}{r_b-r_s} \,.
\end{aligned}
\end{equation}

The coefficients of the recursion for $\Psi_-$ are obtained by sending $\gamma\to-\gamma$.

 \begin{figure}[t]
    \centering
    \includegraphics[width=0.6\textwidth]{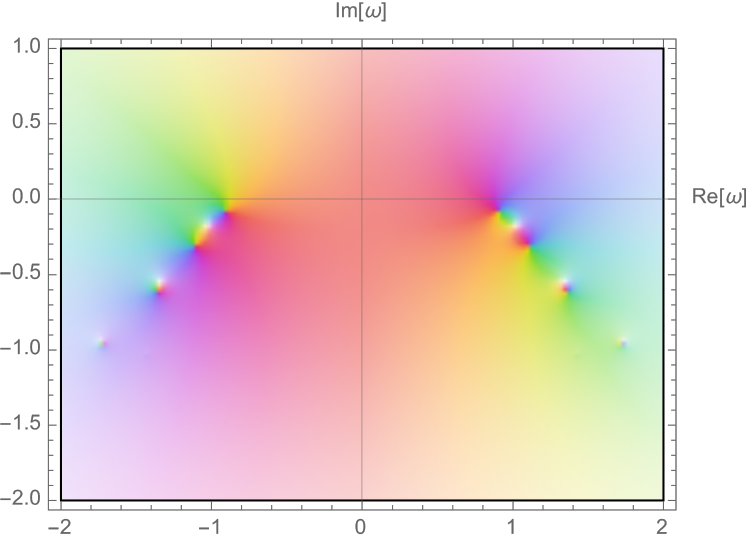}
    \begin{subfigure}[b]{0.35\textwidth}
        \includegraphics[height=0.2\textheight]{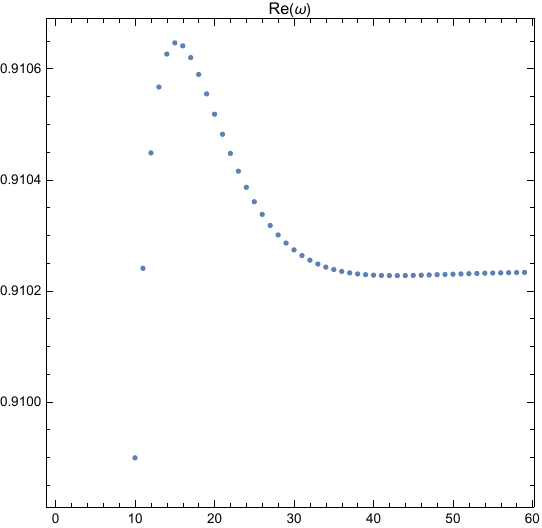}\\
         \includegraphics[height=0.2\textheight]{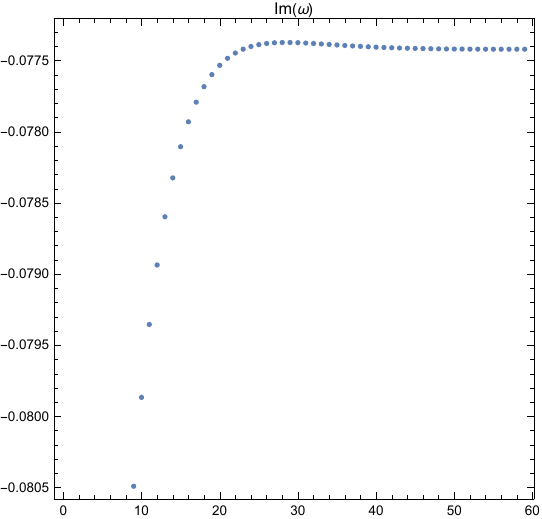}
    \end{subfigure}
    \caption{ Left Panel) QNMs for top star with $r_b=1$, $r_s=0.8$ and $\ell=2$. Right Panels) Convergence of the continuous-fraction method.}
    \label{convTSs2}
\end{figure}

\begin{table}[]
\centering
\begin{tabular}{|c|c|c|}
\hline
$\ell$ & Leaver($n=0$)   & Dir. Int.                  \\ \hline
$2$    & $0.9102 - 0.0774 i$   & $0.9102 - 0.0774 i$   \\ \hline
$3$    & $1.3939 - 0.0477 i$   & $1.3939 - 0.0477 i$   \\ \hline
$4$    & $1.8618 - 0.0286 i$   & $1.8618 - 0.0286 i$   \\ \hline
$5$    & $2.3241 - 0.0164 i$   & $2.3241 - 0.0164 i$   \\ \hline
$8$    & $3.6952 - 0.0019 i$  & $3.6952 - 0.0019 i$  \\ \hline
$10$   & $4.6003 - 0.0003 i$ & $4.6003 - 0.0003 i$ \\ \hline
 \end{tabular}
\quad
 \begin{tabular}{|c|c|}
\hline
$n$ & Leaver   ($\ell=2$)            \\ \hline
$0$  & $0.9102 - 0.0774 i$ \\\hline
$1$ & $1.1201 - 0.3126 i$ \\ \hline
$2$ & $1.3775 - 0.6018 i$  \\ \hline
$3$ & $1.6711 - 0.9105 i$  \\ \hline
$4$ & $1.9868 - 1.2224 i$  \\ \hline
$5$ & $2.3415 - 1.5428 i$  \\ \hline
\end{tabular}
 \caption{QNMs for top star with $r_b=1$, $r_s=0.8$. }
 \label{tabrb1rs08}
\end{table}
Truncating $n$ to some large number  $N$, the recursion relation \eqref{rec} becomes an $N$-dimensional matrix equation $M\cdot c=0$, where $c$ is a vector containing the $c_n$ coefficients entering in the ansatz \eqref{rleaver}. Therefore, a non-trivial solution exists only if the determinant of $M$ vanishes, which provides the equation satisfied by the QNM frequencies. However, we are not going to use the vanishing of the determinant to find the QNMs. Instead, we find more convenient to  first tridiagonalize  $M$,
\begin{equation}
\begin{aligned}
\alpha'_0\,=\,&\alpha_0,\quad \beta'_0=\beta_0\,,\\\nonumber
\alpha'_1\,=\,&\alpha_1,\quad \beta'_1=\beta_1,\quad \gamma'_1=\gamma_1\\\nonumber
\alpha'_n\,=\,&\alpha_n,\quad \beta'_n=\beta_n-\frac{\delta_n}{\gamma'_{n-1}}\alpha'_n,\quad \gamma'_n\,=\,&\gamma_n-\frac{\delta_n}{\gamma'_{n-1}}\beta'_n,\quad \delta'_n=0\,,\quad n\geq2
\end{aligned}
\end{equation}
and then use the fact that the integrability of the three-term recursion boils down to the continuous fraction equation:
\begin{equation}
    \label{fraction Leaver}
\beta'_n +\frac{\alpha'_n\gamma'_n}{\beta'_{n-1}-\frac{\alpha'_{n-2}\gamma'_{n-1}}{\beta'_{n-2}-\dots}}+\frac{\alpha'_n\gamma'_{n+1}}{\beta'_{n+1}-\frac{\alpha'_{n+1}\gamma'_{n+2}}{\beta'_{n+2}-\dots}}=0\, .
\end{equation}
The QNM frequencies can be obtained by solving the above equation for a particular $n$, let us say $n=0$. In practice, the continuous fraction is truncated by keeping $N$ terms, with $N$ a large enough number. In Figure~\ref{convTSs2}  (left) the result for the QNM frequencies is displayed for the $\ell=2$ mode in a top star with $r_s=0.8$, $r_b=1$ with $N=50$. The plots on the right show the convergence of the method, which typically occurs for $N\ge 30$.  
 
 In Table \ref{tabrb1rs08} we display the results for QNM frequencies for the fundamental mode $n=0$ as $\ell$ varies from 2 to 10. 
   The  results  are compared against those obtained from a direct numerical integration of the differential equation, showing an excellent agreement. We omit the results based on a WKB approximation of the solution as the latter fails in reproducing the imaginary part of the QNM frequencies for the low $n$ modes. 
  
Finally, in appendix~\ref{app:qnms} we collect more results for various representative choices of $r_s$, $r_b$, $\ell$.  In all the solutions we have checked, QNM frequencies have negative imaginary parts, suggesting the stability of top stars against metric and electromagnetic odd perturbations.

\subsection{QNM: black string}
\label{sec:qnm_bs}

\begin{figure}[t]
    \centering
    \includegraphics[width=0.5\textwidth]{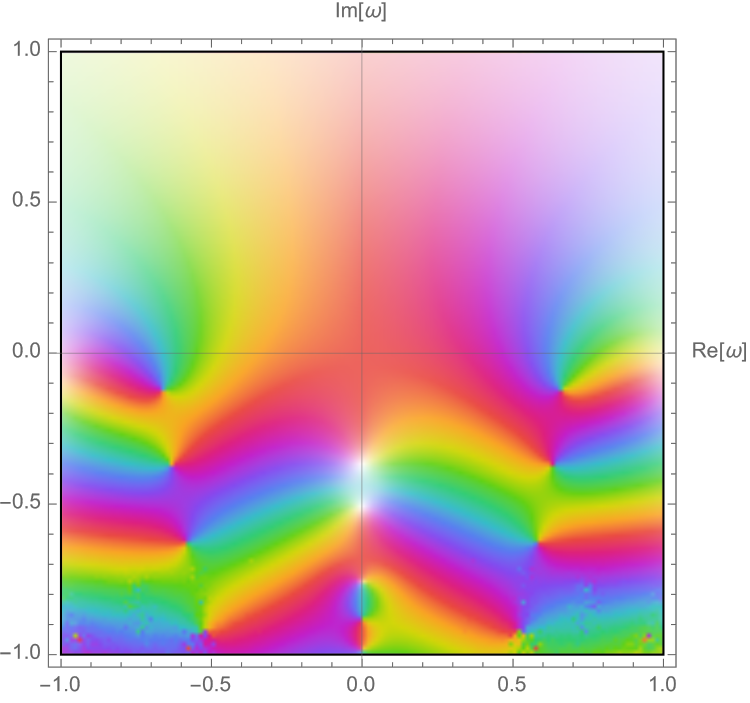} \includegraphics[height=0.33\textheight]{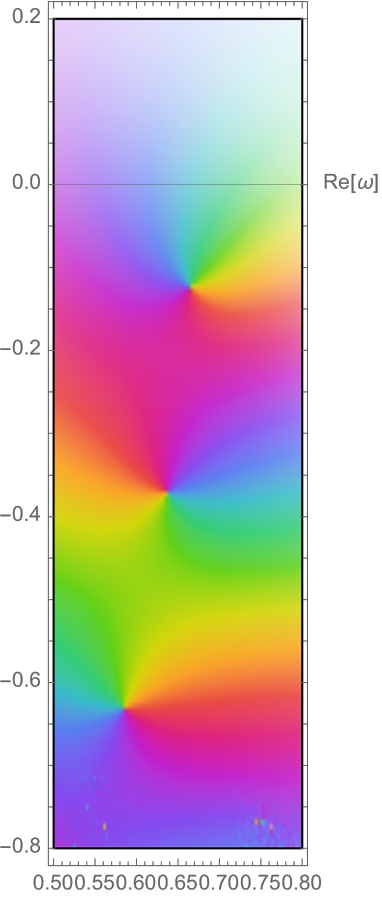}
    \caption{  $R_+$ QNM's for black string with $r_b=0.8$, $r_s=1$, $\ell=2$ }
    \label{figBS}
\end{figure}
 \begin{table}[t]
 \centering
   \begin{tabular}{|c|c|c|c|}
\hline
$\ell$ & WKB               & Leaver      & Dir. Int.          \\ \hline
$2$    & $0.6323 - 0.1239 i$ & $0.6648 - 0.1225 i$ & $0.6648 - 0.1225 i$\\ \hline
$3$    & $1.0608 - 0.1267 i$  & $1.0826 - 0.1263 i$  & $1.083 - 0.1263 i$ \\ \hline
$4$    & $1.4678 - 0.1282 i$  & $1.4839 - 0.1279 i$  & $1.4839 - 0.1279 i$\\ \hline
$5$    & $1.8659 - 0.129 i$   & $1.8787 - 0.1288 i$ & $1.8787 - 0.1288 i$ \\ \hline
$8$    & $3.0402 - 0.1301 i$  & $3.0482 - 0.13 i$  & $3.0482 - 0.13 i$\\ \hline
$10$    & $3.8166 - 0.1304 i$  & $3.8229 - 0.1303 i$  & $3.8229 - 0.1303 i$\\ \hline
\end{tabular}
    \caption{The fundamental $n=0$ QNM for a black string with $r_b=0.8$, $r_s=1$,   }
    \label{tableBS}.
 \end{table}

The analysis for  the black string $r_b< r_s$ follows {\it mutatis mutandis} the same steps than that for topological stars, but now incoming boundary conditions have to be imposed at the horizon $r=r_s$. Therefore, the ansatz now becomes
\begin{equation}
 \Psi(r)=e^{i\omega r}  r^{-{3\over 2}} (r-r_s)^{\lambda_s }(r-r_b)^{\lambda_b}  \sum_{n=0}^\infty c_n\left(\frac{r-r_s}{r-r_b}\right)^n \,, \label{Ransatz}
\end{equation}
with 
\be
\lambda_s = {1\over 2} +{ \omega r_s^{3\over 2} \over \sqrt{r_b-r_s} }    \,, \hspace{1cm}\lambda_b = 1+{{\rm i} \omega   \over 2( r_s-r_b) }   \left( 2 r_s^2-r_b^2-r_b r_s+2 \sqrt{ r_s^3(r_s-r_b) }  \right)\, .
\ee
 Plugging this into (\ref{eqcan})  yields again a four-term recursion relation as in \eqref{rec}, but this time with coefficients given by: 
\begin{equation}
\begin{aligned}
\alpha _n &=-(n+1)^2 r_s+\frac{2 i (n+1) \omega  r_s^2}{\sigma }\,, \nn\\[1mm]
\beta _n &=-\frac{1}{2} r_s \left(-2 \gamma  \sigma ^2+5 \gamma -2 \ell^2-2 \ell+2 n^2 \sigma ^2-6n^2-6 n \sigma ^2+2 n+2 \sigma ^2+1\right) \nn\\
&-\frac{i (\sigma +1) \omega  \left(2 n\sigma ^2-6 n \sigma +12 n+\sigma ^2+5 \sigma -2\right) r_s^2}{2 \sigma}-\frac{(\sigma +1)^3 \omega ^2 r_s^3}{\sigma ^2}\,,\nn\\[1mm]
\gamma _n &=  \frac{1}{2} r_s\left(-2 \gamma  \sigma ^2+5 \gamma +2 \ell^2 \sigma ^2-2 \ell^2+2 \ell \sigma ^2-2 \ell+4 n^2\sigma ^2-6 n^2-16 n \sigma ^2+16 n+14 \sigma ^2-9\right) \,\nn\\
&-\frac{i (\sigma +1)^2\omega  \left(n \sigma ^3-2 n \sigma ^2+6 n \sigma -6 n-2 \sigma ^3+4 \sigma ^2-10\sigma +8\right) r_s^2}{\sigma }+\frac{\left(\sigma ^2-8 \sigma +8\right) (\sigma+1)^4 \omega ^2 r_s^3}{4 \sigma ^2}\,, \nn\\[1mm]
\delta _n &=  -\frac{i (2 n-5) (\sigma -2) (\sigma-1) (\sigma +1)^3 \omega  r_s^2}{2 \sigma }-(n-3) (n-2) (\sigma -1) (\sigma +1)r_s \nn\\
& +\frac{(\sigma -2)^2 (\sigma -1) (\sigma +1)^5 \omega ^2 r_s^3}{4 \sigma
   ^2} \,,
\end{aligned}
\end{equation}
with
\be
\sigma=\sqrt{ 1-{r_b\over r_s}}\, .
\ee
Figure~\ref{figBS} shows some QNM frequencies (different overtones) $r_s=1$, $r_b=0.8$ and $\ell=2$, again for the $\Psi_{+}$ perturbation. Those for $n=0$ with $\ell$ varying from $2$ to $10$ are displayed in Table~\ref{tableBS}, together with a comparison against the WKB and direct integration methods. As we can see, we find an pretty good agreement. Finally, other representative examples are collected in the Appendix~\ref{app:qnms}. For $r_b=0$ we reproduce the results for Schwarzschild black holes, as expected. 

As in the analysis of the topological star, we find that for all choices of $r_s$, $r_b$, $\ell$  we have checked the imaginary parts of the QNM frequencies of black strings are always negative, strongly suggesting the stability of these solutions against this type of perturbations.


\section{Gregory-Laflamme instabilities}
\label{sec:GL}

\begin{figure}[t]
    \centering
    {\includegraphics[width=0.45\textwidth]{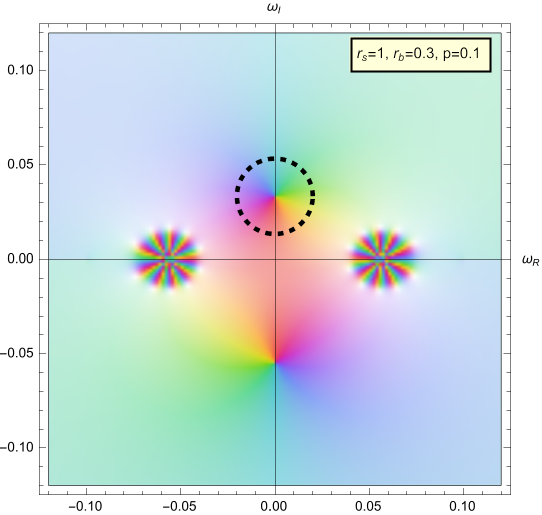}}
    {\includegraphics[width=0.45\textwidth]{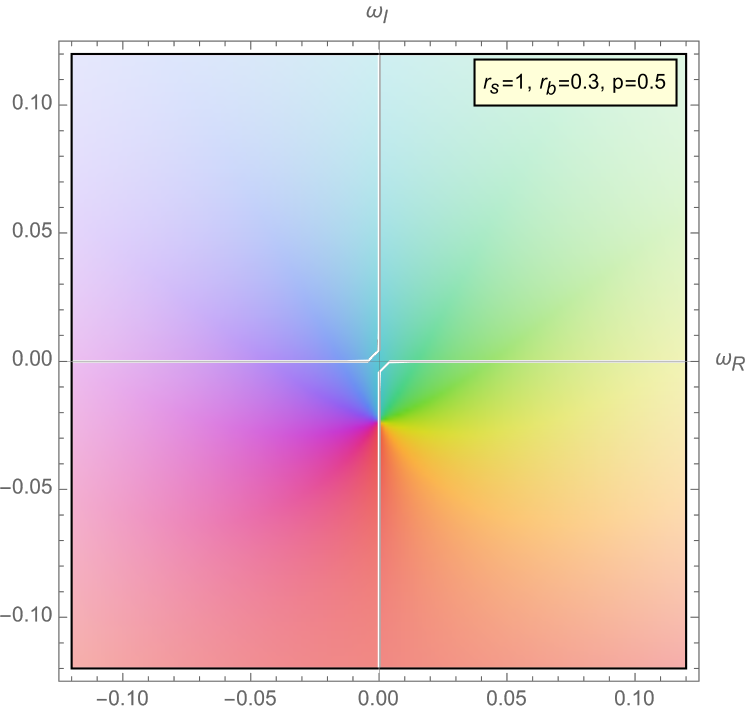}}\\
 { \includegraphics[width=0.45\textwidth]{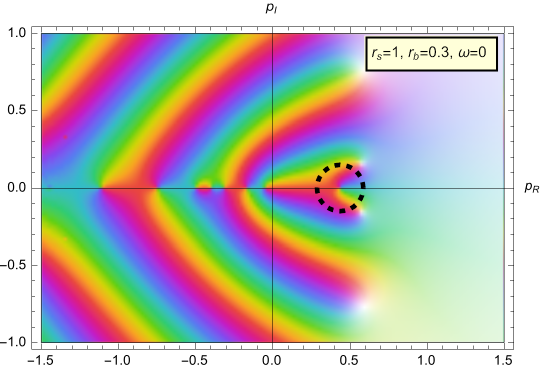}}  {\includegraphics[width=0.45\textwidth]{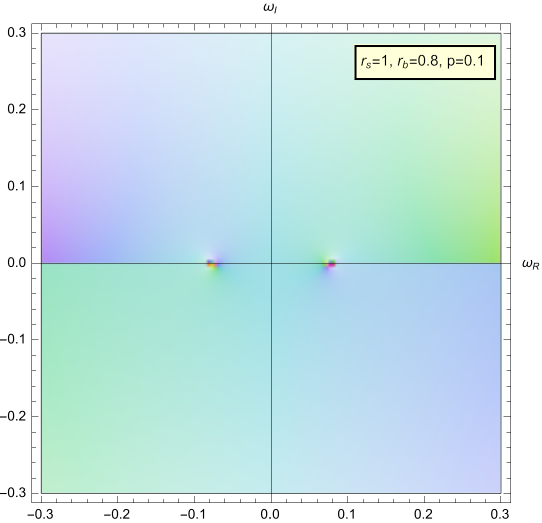}}  
    \caption{Black string unstable modes for various parameters; the frequency of the unstable mode is at $\omega=0.0332837 i$ (Top left), $p=0.435943$ (Bottom left).}
    \label{fig:GL1}
\end{figure}

Black strings and branes in dimensions higher than four typically exhibit a classical instability, known as the Gregory-Laflamme (GL) instability \cite{Gregory:1994bj}. A characteristic feature of the GL perturbations is that they have a momentum, $p$, along the internal directions.  Typically, the instability appears when $p$ is smaller than a certain threshold value, $p_{\star}$, while the system is stable under perturbations with $p>p_{\star}$. This suggests that the mode with $p=p_{\star}$ (which is referred to as the \emph{threshold unstable mode}) is time independent. 

Based on the existence of a threshold time-independent mode, the domain of GL stability of black strings was determined in \cite{Miyamoto:2007mh} to be $r_s<2 r_b$. Furthermore, \cite{Bah:2021irr,Stotyn:2011tv} used the double Wick rotation symmetry that exchanges black strings with topological stars, $r_s\leftrightarrow r_b$  and $\omega \leftrightarrow {\rm i} p$  to argue that top stars are stable when $r_b < 2 r_s$.\footnote{In the analysis of \cite{Bah:2021irr},  the role of the threshold mode is played by a mode with $p=0$ and $\omega = {\rm i} p_{\star}$.} We also know that when $r_b$ is fixed and $r_s=0$, the top star becomes Euclidean Schwarzschild times time, and this solution suffers from a Gross-Perry-Yaffe instability \cite{Gross:1982cv, Witten:1981gj}. Hence, we expect the instability of top stars with $0< r_s <  r_b/2$ to be of the same type \cite{Bah:2021irr,Stotyn:2011tv}.

 Here we review the arguments of \cite{Miyamoto:2007mh,Bah:2021irr}, adapting them to the more general perturbations where both $p$ and $\omega$ are non-vanishing. We then apply the Leaver method to compute the QNM frequencies and show directly the existence of modes with positive imaginary parts outside of the stability domains.
 
Let us consider the following even perturbation of the metric and two-form with $\ell=0$ (hence, spherically-symmetric),
\begin{equation}
\begin{aligned}
h_{\mu\nu}\diff {x}^\mu \diff {x}^{\nu}\,=\,& e^{{\rm i} p y-{\rm i} \omega t} \left[  h_{1}(r) \diff r^2{+}r^2 k(r) \left(\diff \theta^2{+}\sin^2\theta\,\diff \phi^2\right) {+}  2\,h_{2}(r) \,\diff r \,\diff y {+} 2\, h_{3}(r) \,\diff r \,\diff t \right] 
\,,\\[1mm]
c_{\mu\nu}\diff {x}^\mu \diff {x}^{\nu}\,=\,& 2 e^{{\rm i} p y-{\rm i} \omega t} c(r) \,\diff t\wedge \diff y\,.
\end{aligned}
\end{equation}
One can check that for any choice of $p$ and $\omega$ this is not pure gauge:  $h_{\mu\nu}$ cannot be written as $\nabla_{(\mu}\zeta_{\nu)}$. The field equations \eqref{eq:einstein} and \eqref{eq:Maxwell} allow us to solve for $c$, $h_{1}$, $h_2$ and $h_{3}$ in terms of $k$, and we find that the latter satisfies the second-order differential equation
\begin{equation}
\label{eq:k}
{d\over dr} \left[ A(r)  k'(r)  \right] - B(r) k(r) =0\,,
\end{equation}
with
\begin{equation}
\begin{aligned}
A(r) &=\frac{\left(r{-}r_b\right) \left(r{-}r_s\right)}{ {\cal W}(r)^2}\,, \label{abr}\\[1mm]
B(r)& = \frac{ r^3 {\cal W}(r)  \left[p^2 \left(r{-}r_s\right){-}\omega ^2 \left(r{-}r_b\right)\right]+2 \left(r{-}r_b\right) \left(r{-}r_s\right) \left[p^2 \left(2 r_b{-}r_s\right){-}\omega ^2
   \left(2 r_s{-}r_b\right)\right]}{{\cal W}(r)^3 \left(r{-}r_b\right) \left(r{-}r_s\right)}\,,
\end{aligned}
\end{equation}
 and
 \begin{equation}
{\cal W}(r)=p^2 \left(4 r-3 r_s\right)-\omega ^2 \left(4 r-3 r_b\right)\, .
 \end{equation}
 
 \begin{figure}[t]
    \centering
   {\includegraphics[width=0.45\textwidth]{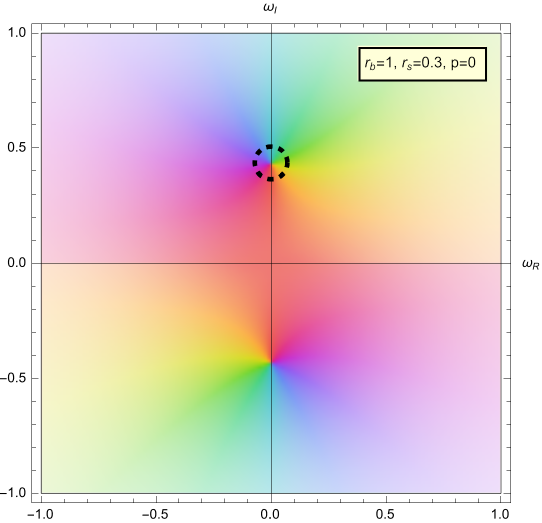}}
    \hfill
   {\includegraphics[width=0.45\textwidth]{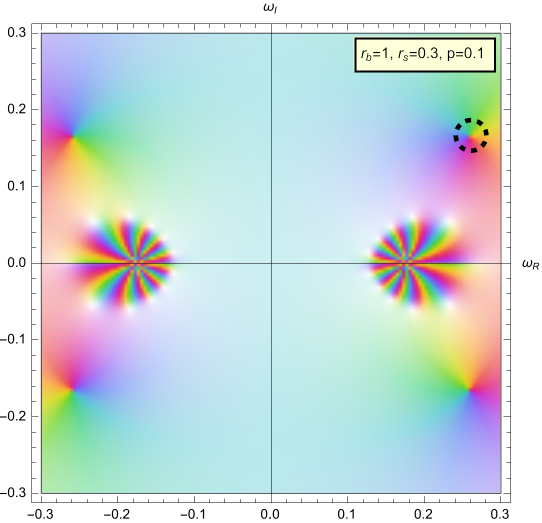}}
    \\
   {\includegraphics[width=0.45\textwidth]{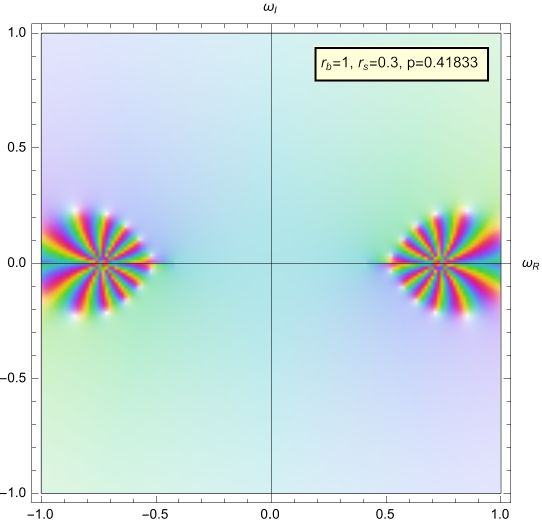}}
    \hfill
    {\includegraphics[width=0.45\textwidth]{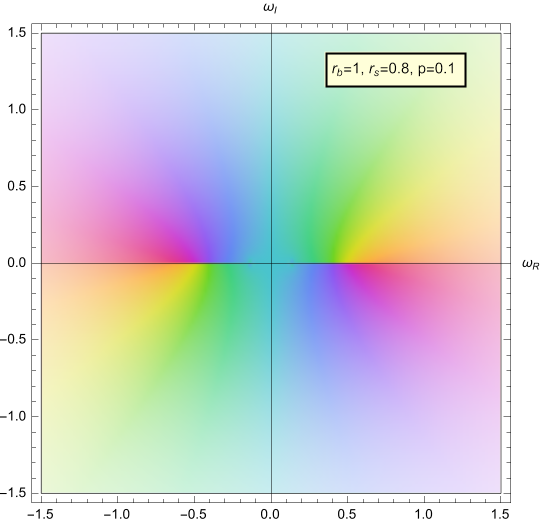}}
    \caption{GL unstable modes for the topological star for various parameters: the frequencies are $\omega=0.435929 i$ (top left) and $\omega=0.260928 - 0.166396 i$ (top right).}
    \label{fig:GL2}
\end{figure}

We look for solutions to \eqref{eq:k} satisfying the boundary conditions,
 \begin{equation}\label{bcGL}
k(r)\underset{r\to r_0}{\sim} (r-r_0)^{\lambda_0} \,, \hspace{2cm} k(r)\underset{r\to \infty}{\sim} e^{ {\rm i} r \sqrt{ \omega^2-p^2}}\, , 
 \end{equation}
where $r_0={\rm {max}}\left[r_s, r_b\right]$ and $\lambda_0$ is determined by solving the equation near $r=r_0$ and imposing incoming (for the black string) or regular (for the top star) boundary conditions at $r=r_0$. The GL modes correspond to solutions of (\ref{eq:k}) satisfying (\ref{bcGL}), and exhibiting an exponential decay at infinity \cite{Gregory:1994bj}: 
\begin{align}
    {\rm Re}\left( {\rm i} \sqrt{ \omega^2-p^2}\right) <0\, . \label{cond}
\end{align}
In order to review the arguments presented in \cite{Miyamoto:2007mh,Bah:2021irr} in a unified way, let us first consider perturbations where $p$ is real and $\omega$ is purely imaginary and  \eqref{cond}  is automatically satisfied. 
Multiplying (\ref{eq:k}) by $k(r)$, integrating it over the domain $ [r_0 , \infty]$,  and integrating by parts, we obtain:
\begin{equation}
\label{eq:k2}
\int^{\infty}_{r_0} \diff r \left[A(r) \left(k'(r)\right)^2+ B(r)   k(r)^2\right]=\int^{\infty}_{r_0} d\left[ A(r) k(r) k'(r)\right] =0\, .
\end{equation}
 The vanishing of the right hand side follows from the fact that $A(r)$ vanishes at $r_0$ and $k(r)$ vanishes at infinity. In addition if $p$ is real and $\omega$ purely imaginary, both $A(r)$ and $B(r)$ are real and positive in the domain:
\begin{equation}
{\rm{top~star:}}\,\,\, r_s< r_b < 2 r_s \,, \hspace{1cm}{\rm{black~string:}}\,\,\,r_b< r_s < 2 r_b\,.
\end{equation}
Thus, the only (real) solution of (\ref{eq:k2}) is the trivial one, $k=0$. The geometry is therefore stable in this domain.
  
Given this, we provide further evidence applying again Leaver's method. Since the procedure is analogous to the one described in the previous section, we shall ignore the technical details and present directly the results in Figures~\ref{fig:GL1} and \ref{fig:GL2}, where we plot the argument of the QNM eigenvalue equation for various choices of $r_s$ and $r_b$. The QNM frequencies show up as peaks on these plots. 

A summary of our findings is:

\begin{itemize}
\item Black string (Figure~\ref{fig:GL1}): We find unstable modes only in the regime $r_s>2r_b$ for $p<p_{\star}\approx 0.435943$ in agreement with 
expectations in \cite{Miyamoto:2007mh}.  

\item Top star (Figure~\ref{fig:GL2}):  We find unstable modes for $r_b>2r_s$ and $p$ small enough. However when $p$ is finite, it cannot be arbitrarily small since regularity of the geometry 
requires the quantization condition:
\begin{equation}
p=\frac{n_y}{R_y}=n_y \,\frac{{\sqrt{r_b-r_s}}}{2 r_b^{3/2}}\,, \hspace{1cm} n_y \in {\mathbb Z}\,,
\end{equation}
We find that all modes with $n_y \ge 1$ are stable, therefore only the  $n_y=0$ mode leads to a potential instability. 

\end{itemize}


\section{Conclusions}
\label{sec:conclusions}

We have calculated the QNM frequencies associated to odd perturbations of topological stars and magnetized black strings in five-dimensional Einstein-Maxwell theory. 
Our analysis of the QNM spectrum mainly relies on Leaver's method \cite{Leaver:1985ax}, which has been suitably generalized in order to apply it to an ODE with five Fuchsian singularities, reducing the 4-term recursion relation to a 3-term one via a tri-diagonalization of the eigenvalue matrix.

We have confirmed the expectations of \cite{Bah:2021irr, Stotyn:2011tv} which suggest classical stability inside the domain
\begin{equation}
{\rm{top~star:}}\,\,\, r_s< r_b < 2 r_s \,, \hspace{1cm}{\rm{black~string:}}\,\,\,r_b< r_s < 2 r_b\,,
\end{equation}
as we have found that all the perturbations we have considered inside the above domains decay exponentially in time.  Outside of these domains we have verified that topological stars suffer from a Gross-Perry-Yaffe-type instability \cite{Gross:1982cv}, whereas magnetic black strings suffer from the usual Gregory-Laflamme instability \cite{Miyamoto:2007mh}.  Such instabilities reflect themselves on the existence of QNMs with positive imaginary part which are associated to even perturbations with $\ell=0$.

There are several obvious next steps to our investigation. The first is to study the possible gravitational-wave signatures that may distinguish top stars from black holes: multipolar structure\cite{Mayerson:2020tpn,Bianchi:2020des}, echoes \cite{Ikeda:2021uvc} and tidal effects during the inspiral phase of black-hole merger \cite{Fucito:2023afe}. 

The next is to consider the even perturbations, and try to solve the underlying system of ODE's numerically. This will involve a shooting problem in several functions of one variable. Systems of similar complexity have been solved by shooting in other circumstances \cite{Ganchev:2021pgs, Buchel:2000ch}, so we believe this problem is within reach.

Another extension of our calculation is to the running-Bolt solutions \cite{Bena:2009qv}, which are also obtained by magnetizing the bolt of Euclidean Schwarzschild times time in five dimensions. For these solutions we also expect a Gross-Perry-Yaffe-type instability for small magnetic fluxes, and possibly a stable solution for larger fluxes \cite{Avery}. Another cohomogeneity-one solution with fluxes that can be studied using our method is the magnetized Atiyah-Hitchin solution \cite{Atiyah:1985dv,Bena:2007ju}, which is the M-theory uplift of Type-IIA Orientifold 6-planes with fluxes.

Our calculation serves as a first step towards and a benchmark in the determination of the stability or lack thereof for more generic non-extremal topological stars that have a fluxed bolt. The most generic such solutions are cohomogeneity-two \cite{Bah:2021owp}, and hence determining the quasinormal frequencies will require solving PDE's, using methods similar to those of \cite{Dias:2009iu,Dias:2014eua,Santos:2015iua}. These calculations should reduce in certain limits to the calculations we do here, and hence our calculations should serve as a useful benchmark for these more complicated calculations.



\section*{Acknowledgments:}We would like to thank Massimo Bianchi, Pablo A.~Cano, Giuseppe Dibitetto, Alexandru Dima, Francesco Fucito, Pierre Heidmann, Marco Melis, Paolo Pani and David Pereniguez for stimulating discussions and exchanges. The work of  IB is supported in part by the ERC Grants 787320 - QBH Structure and 772408 - Stringlandscape.
The work of GDR, JFM and AR is supported by the MIUR-PRIN contract 2020KR4KN2 - String Theory as a bridge between Gauge Theories and Quantum Gravity and by the INFN Section of Rome ``Tor Vergata''.

\begin{appendix}

\section{QNMs}
\label{app:qnms}
 In this appendix we collect some tables displaying QNM frequencies corresponding to the perturbations $\Psi_{+}$ (left) and $\Psi_{-}$ (right) for various choices of $r_s$, $r_b$ and $\ell$.

 \begin{itemize}
 
 \item{ $r_s=0.8$, $r_b=1$:
 
\begin{table}[h]
\centering
\begin{tabular}{|c|c|c|}
\hline
$\ell$ & Leaver  ($n=0$)   & Dir. Int.    ($n=0$)                \\ \hline
$2$    & $0.9102 - 0.0774 i$   & $0.9102 - 0.0774 i$   \\ \hline
$3$    & $1.3939 - 0.0477 i$   & $1.3939 - 0.0477 i$   \\ \hline
$4$    & $1.8618 - 0.0286 i$   & $1.8618 - 0.0286 i$   \\ \hline
$5$    & $2.3241 - 0.0164 i$   & $2.3241 - 0.0164 i$   \\ \hline
$6$    & $2.7833 - 0.0087 i$  & $2.7833 - 0.0087 i$  \\ \hline
$7$    & $3.2403 - 0.0043 i$  & $3.2403 - 0.0043 i$  \\ \hline
$8$    & $3.6952 - 0.0019 i$  & $3.6952 - 0.0019 i$  \\ \hline
$9$    & $4.1484 - 0.0008 i$  & $4.1484 - 0.0008 i$  \\ \hline
$10$   & $4.6003 - 0.0003 i$ & $4.6003 - 0.0003 i$ \\ \hline
\end{tabular}
\quad
\begin{tabular}{|c|c|c|}
\hline
$\ell$ & Leaver   ($n=0$)  & Dir. Int      ($n=0$)              \\ \hline
$2$    & $1.5803 - 0.1064 i$   & $1.5803 - 0.1064 i$   \\ \hline
$3$    & $2.0518 - 0.0694 i$   & $2.0518 - 0.0694 i$   \\ \hline
$4$    & $2.5181 - 0.0454 i$   & $2.5181 - 0.0454 i$   \\ \hline
$5$    & $2.9815 - 0.0288 i$  & $2.9815 - 0.0288 i$   \\ \hline
$6$    & $3.4429 - 0.0174 i$  & $3.4429 - 0.0174 i$  \\ \hline
$7$    & $3.9025 - 0.0098 i$  & $3.9025 - 0.0098 i$  \\ \hline
$8$    & $4.3604 - 0.0051 i$  & $4.3604 - 0.0051 i$  \\ \hline
$9$    & $4.8166 - 0.0024 i$  & $4.8166 - 0.0024 i$  \\ \hline
$10$   & $5.2711 - 0.001 i$ & $5.2711 - 0.001 i$ \\ \hline
\end{tabular}
\end{table}
\begin{table}[h]
\centering
\begin{tabular}{|c|c|}
\hline
$n$ & Leaver     ($\ell=2$)           \\ \hline
$0$  & $0.9102 - 0.0774 i$ \\\hline
$1$ & $1.1201 - 0.3126 i$ \\ \hline
$2$ & $1.3775 - 0.6018 i$  \\ \hline
$3$ & $1.6711 - 0.9105 i$  \\ \hline
$4$ & $1.9868 - 1.2224 i$  \\ \hline
$5$ & $2.3415 - 1.5428 i$  \\ \hline
\end{tabular}
\quad
\begin{tabular}{|c|c|}
\hline
$n$ &  Leaver    ($\ell=2$)           \\ \hline
$0$  &  $1.5803 - 0.1064 i$  \\\hline
$1$  & $1.7748 - 0.345 i$ \\ \hline
$2$   & $2.0085 - 0.6302 i$  \\ \hline
$3$  & $2.274 - 0.9405 i$  \\ \hline
$4$  & $2.5645 - 1.2634 i$  \\ \hline
$5$  & $2.8598 - 1.5919 i$  \\ \hline
\end{tabular}
\end{table}
\begin{table}[h]
\centering
\begin{tabular}{|c|c|c|}
\hline
$n$ & Leaver ($\ell=10$) & Dir. Int. ($\ell=10$)            \\ \hline
$0$ & $4.6003 - 0.0003 i$  &  $4.6003 - 0.0003 i$  \\\hline
$1$ & $4.7872 - 0.0321 i$ & $4.7872 - 0.0321 i$\\ \hline
$2$ & $4.9538 - 0.1546 i$ & $4.9538 - 0.1546 i$\\ \hline
$3$ & $5.139 - 0.3201 i$ & $-$ \\ \hline
$4$ & $5.3355 - 0.5135 i$ & $-$ \\ \hline
$5$ & $5.5417 - 0.729 i$ & $-$ \\ \hline
\end{tabular}
\quad
\begin{tabular}{|c|c|c|}
\hline
$n$ & Leaver  ($\ell=10$) & Dir. Int. ($\ell=10$)          \\ \hline
$0$ & $5.2711 - 0.001 i$ & $5.2711 - 0.001 i$ \\ \hline
$1$ & $5.4367 - 0.0486 i$ & $5.4368 - 0.0486 i$ \\ \hline
$2$ & $5.5997 - 0.1799 i$  & $-$  \\ \hline
$3$  & $5.7793 - 0.348 i$ & $-$  \\ \hline
$4$  & $5.9699 - 0.5423 i$ & $-$  \\\hline
$5$  & $6.1702 - 0.7575 i$  &$-$  \\\hline
\end{tabular}
 \end{table}
}

\newpage
 \item{ $r_s=0$, $r_b=1$:

\begin{table}[h]
\centering
\begin{tabular}{|c|c|}
\hline
$\ell$ & Leaver     ($n=0$)   \\ \hline
$2$    & $1.8517 - 1.2691 i$     \\ \hline
$3$    & $3.0513 - 1.3549 i$   \\ \hline
$4$    & $4.157 - 1.3815 i$   \\ \hline
$5$    & $5.222 - 1.3933 i$   \\ \hline
$6$    & $6.266 - 1.3996 i$   \\ \hline
$7$    & $7.2979 - 1.4034 i$     \\ \hline
$8$    & $8.3221 - 1.4059 i$   \\ \hline
$9$    & $9.3411 - 1.4076 i$   \\ \hline
$10$   & $10.3564 - 1.4088 i$   \\ \hline
\end{tabular}
\quad
\begin{tabular}{|c|c|}
\hline
$\ell$ & Leaver    ($n=0$)            \\ \hline
$2$    & $2.2995 - 1.4362 i$     \\ \hline
$3$    & $3.3569 - 1.425 i$   \\ \hline
$4$    & $4.3888 - 1.4206 i$   \\ \hline
$5$    & $5.409 - 1.4185 i$   \\ \hline
$6$    & $6.423 - 1.4173 i$   \\ \hline
$7$    & $7.4333 - 1.4165 i$     \\ \hline
$8$    & $8.4412 - 1.416 i$   \\ \hline
$9$    & $9.4474 - 1.4156 i$   \\ \hline
$10$   & $10.4524 - 1.4154 i$   \\ \hline
\end{tabular}
 \end{table}
\begin{table}[h]
\centering
\begin{tabular}{|c|c|}
\hline
$n$  & Leaver    ($\ell=2$)           \\ \hline
$0$  & $1.8517 - 1.2691 i$ \\ \hline
$1$  & $1.4597 - 2.7725 i$ \\ \hline
$2$   & $1.0743 - 4.6179 i$  \\ \hline
$3$    & $0.8437 - 6.6137 i$  \\ \hline
$4$    &   $0.7107 - 8.6372 i$     \\ \hline
$5$   &   $0.6269 - 10.6623 i$\\ \hline
\end{tabular}
\quad
\begin{tabular}{|c|c|}
\hline
$n$  & Leaver    ($\ell=2$)               \\ \hline
$0$  &  $2.2995 - 1.4362 i$  \\\hline
$1$  & $1.9111 - 3.0645 i$ \\ \hline
$2$   & $1.5047 - 4.9713 i$  \\ \hline
$3$    & $1.2273 - 7.0143 i$  \\ \hline
$4$    &   $1.0492 - 9.0811 i$     \\ \hline
$5$   &   $0.927 - 11.1432 i$\\ \hline
\end{tabular}
\end{table}
\begin{table}[h]
\centering
\quad
\begin{tabular}{|c|c|}
\hline
$n$  & Leaver     ($\ell=10$)             \\ \hline
$0$  &  $10.3564 - 1.4088 i$  \\\hline
$1$  & $10.2491 - 2.828 i$ \\ \hline
$2$   & $10.0721 - 4.2681 i$  \\ \hline
$3$   & $9.8284 - 5.7402 i$  \\ \hline
$4$    & $9.5235 - 7.2556 i$  \\ \hline
$5$    &   $9.1654 - 8.8256 i$     \\ \hline
\end{tabular}
\begin{tabular}{|c|c|}
\hline
$n$  & Leaver      ($\ell=10$)           \\ \hline
$0$  &  $10.4524 - 1.4154 i$   \\\hline
$1$  & $10.3457 - 2.841 i$ \\ \hline
$2$   & $10.1698 - 4.2876 i$  \\ \hline
$3$    & $9.9277 - 5.766 i$  \\ \hline
$4$    &   $9.6249 - 7.2876 i$     \\ \hline
$5$   &   $9.2694 - 8.8632 i$\\ \hline
\end{tabular}
\end{table}
}

\newpage 

\item{ $r_b=0.8$, $r_s=1$:

\begin{table}[h]
\centering
\begin{tabular}{|c|c|c|c|}
\hline
$\ell$ & WKB      ($n=0$)            & Leaver    ($n=0$)     & Dir. Int.     ($n=0$)        \\ \hline
$2$    & $0.6323 - 0.1239 i$ & $0.6648 - 0.1225 i$ & $0.6648 - 0.1225 i$\\ \hline
$3$    & $1.0608 - 0.1267 i$  & $1.0826 - 0.1263 i$  & $1.083 - 0.1263 i$ \\ \hline
$4$    & $1.4678 - 0.1282 i$  & $1.4839 - 0.1279 i$  & $1.4839 - 0.1279 i$\\ \hline
$5$    & $1.8659 - 0.129 i$   & $1.8787 - 0.1288 i$ & $1.8787 - 0.1288 i$ \\ \hline
$6$    & $2.2595 - 0.1295 i$  & $2.2702 - 0.1294 i$  & $2.2702 - 0.1294 i$\\ \hline
$7$    & $2.6507 - 0.1299 i$  & $2.6598 - 0.1297 i$  & $2.6598 - 0.1297 i$\\ \hline
$8$    & $3.0402 - 0.1301 i$  & $3.0482 - 0.13 i$  & $3.0482 - 0.13 i$\\ \hline
$9$    & $3.4288 - 0.1303 i$  & $3.4358 - 0.1302 i$  & $3.4358 - 0.1302 i$\\ \hline
$10$    & $3.8166 - 0.1304 i$  & $3.8229 - 0.1303 i$  & $3.8229 - 0.1303 i$\\ \hline
\end{tabular}
\begin{tabular}{|c|c|c|c|}
\hline
$\ell$ & WKB    ($n=0$)           & Leaver   ($n=0$)      & Dir. Int.   ($n=0$)          \\ \hline
$2$    & $1.0695 - 0.1349 i$ & $1.0925 - 0.1335 i$ & $1.0925 - 0.1335 i$\\ \hline
$3$    & $1.4785 - 0.1338 i$  & $1.4952 - 0.1332 i$  & $1.4952 - 0.1332 i$ \\ \hline
$4$    & $1.8773 - 0.1333 i$  & $1.8905 - 0.1329 i$  & $1.8905 - 0.1329 i$\\ \hline
$5$    & $2.2714 - 0.133 i$   & $2.2822 - 0.1327 i$ & $2.2822 - 0.1327 i$ \\ \hline
$6$    & $2.6627 - 0.1328 i$  & $2.672 - 0.1326 i$  & $2.672 - 0.1326 i$\\ \hline
$7$    & $3.0524 - 0.1326 i$  & $3.0605 - 0.1325 i$  & $3.0605 - 0.1325 i$\\ \hline
$8$    & $3.441 - 0.1325 i$  & $3.4482 - 0.1324 i$  & $3.4482 - 0.1324 i$\\ \hline
$9$    & $3.8289 - 0.1324 i$  & $3.8353 - 0.1323 i$  & $3.8353 - 0.1323 i$\\ \hline
$10$    & $4.2162 - 0.1323 i$  & $4.2221 - 0.1322 i$  & $4.2221 - 0.1322 i$\\ \hline
\end{tabular}
\end{table}

 \begin{table}[h]
\centering
\begin{tabular}{|c|c|c|}
\hline
$n$ & WKB    ($\ell=2$)           & Leaver      ($\ell=2$)    \\\hline 
$0$    & $0.6323 - 0.1239 i$  & $0.6648 - 0.1225 i$ \\ \hline
$1$    & $0.6323 - 0.3716 i$  & $0.6369 - 0.372 i$    \\ \hline
$2$    & $0.6323 - 0.6194 i$ & $0.5891 - 0.6385 i$   \\ \hline
$3$    & $0.6323 - 0.8671 i$  & $0.5083 - 0.9566 i$     \\ \hline
$4$    & $0.6323 - 1.1149 i$  & $0.4527 - 1.2544 i$     \\ \hline
$5$    & $0.6323 - 1.3626 i$  & $0.3972 - 1.603 i$     \\ \hline
\end{tabular}
\quad
\begin{tabular}{|c|c|c|}
\hline
$n$ & WKB       ($\ell=2$)             & Leaver      ($\ell=2$)               \\ \hline
$0$    & $1.0695 - 0.1349 i$  & $1.0925 - 0.1335 i$  \\\hline
$1$    & $1.0695 - 0.4047 i$  & $1.0785 - 0.4024 i$   \\ \hline
$2$    & $1.0695 - 0.6745 i$ & $1.0518 - 0.6752 i$   \\ \hline
$3$    & $1.0695 - 0.9443 i$  & $1.0226 - 0.9586 i$    \\\hline
$4$    & $1.0695 - 1.2141 i$  & $0.9858 - 1.2474 i$    \\\hline
$5$    & $1.0695 - 1.484 i$  & $0.9531 - 1.531 i$    \\\hline
\end{tabular}
\end{table}

}

\end{itemize}

\newpage

\end{appendix}

\bibliographystyle{JHEP}
\bibliography{references}
\end{document}